# Low cost Ge/Si virtual substrate through dislocation trapping by nanovoids

Youcef A. Bioud, Abderraouf Boucherif, Etienne Paradis,
Ali Soltani, Dominique Drouin and Richard Arès

Laboratoire Nanotechnologies Nanosystèmes (LN2) - CNRS UMI-3463, Institut Interdisciplinaire d'Innovation Technologique (3IT), Université de Sherbrooke, 3000 Boulevard Université, Sherbrooke, J1K OA5, Québec, Canada
E-mail: Richard.Arès@USherbrooke.ca

**Abstract**
A low-cost method to reduce the threading dislocations density (TDD) in relaxed germanium (Ge) epilayers grown on silicon (Si) substrates is presented. Ge/Si substrate was treated with post epitaxial process to create a region with a high density of nanovoids in Ge layer which act as a barrier for threading dislocations propagation.

## 1. Introduction

Multi-junction III-V concentrator cells have the most important energy conversion efficiency, competing with any other solar cell technology [1]. Germanium appears to be the material of choice for the bottom cell to cover the IR part of the solar spectrum. A high efficiency of 40.7% was measured for a metamorphic three junctions GaInP/GaInAs/Ge cell at 240 suns [2]. This architecture has achieved system efficiency of 30% [3]. A road map based on the integration of Ge on Si wafers can be a solution to replace bulk Ge substrates and would result in a significant cost reduction. However, due to the important 4.2% lattice mismatch between Si and Ge, strain is introduced in epitaxial Ge layers on Si which leads to the generation of a high density of misfit dislocations. Dislocations act as recombination centers for carriers and strongly degrade the solar cell performance. Several techniques of dislocation reduction were reported in recent years such as graded SiGe buffer layers, cyclic annealing, aspect-ratio-trapping (ART), selective area depositions (SAD) and deep substrate patterning (DSP) [4]. However, the treading dislocations density (TDD) in the Ge layer remains higher than $10^6$ cm$^{-1}$, reducing its usefulness for many device applications. In this talk, we will discuss a new strategy for making cost effective Ge/Si virtual substrate which is based on a simple and cost effective technique that uses nanovoids as a barrier for dislocation propagation. Then, we will present detailed optimization of the porosification and the annealing of Ge on Si and study the effect of different process parameters on TDD and on surface roughness.

## 2. Results and Discussion
*Approach*

As sketched in Fig 1, Ge/Si substrate was treated with post epitaxial process to create a region with a high density of nanovoids within the Ge layer on Si is expected to act as a barrier for threading dislocations propagation. Nanovoids are formed by electrochemical porosification followed by thermal annealing. The process is completed by CMP (chemical mechanical planarization) step to obtain an epi-ready virtual substrate.

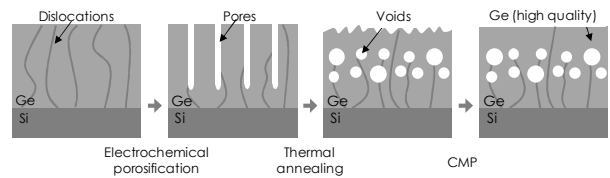

Fig. 1 Schematic illustration of dislocation trapping by nanovoids approach.

*Electrochemical etching of dislocated Ge/Si layer*

The porous morphologies obtained from the electrochemical etching of Ge/Si layer are quite different from that of the bulk Ge. For bulk Ge, the layer structure shows uniformly distributed mesopores with controllable nanostructure and size [5],[8]. However, for dislocated Ge/Si, non-uniformly distributed mesopores are obtained (see Fig 2a and 4b). The porous layer consists of three zones: (i) a thin PGe layer formed at the surface, (ii) a thick PGe layer with weakly branched pores that trend to follow threading dislocation cores and (iii) porous Si layer obtained by HF crossing dislocations.

Calibration of etching rates has been done by varying the current density in the electrochemical process. Fig 2b illustrates the dependence of the etching rate on the current density of bulk and dislocated Ge. The etch rate ratio between dislocated and bulk regions shows very high values up to 6. The preferential etching via dislocations can be explained by their high electrical conduction and their role in the transport properties with the electrolyte.

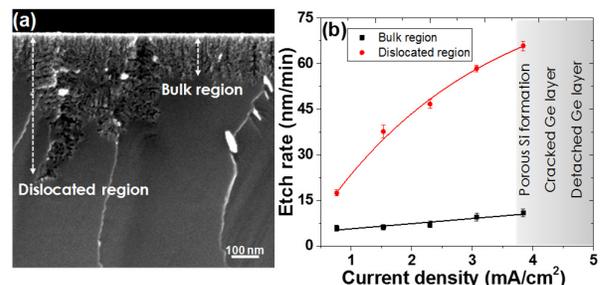

Fig. 2 Cross-section SEM images of porous Ge/Si layer formed at 1.5mA/cm$^2$ (a) and (b) etch rates versus current density of dislocated and bulk regions in Ge/Si layer.

*Nanovoid layer formation*

The morphological change of porous Ge during annealing is based on thermally activated jumps of surface atoms to stable positions following to the mass transport mechanism, for which atoms have a maximum of covalent bonds until their outer energy level is full. In terms of geometry, atoms diffuse from the larger curvature surface toward the lower curvature surface down a gradient potential [9]. Therefore, the pores in the volume are enlarged and the pores on the surface are closed. This reveals the formation of a crystalline region in the top surface and a voided region in the volume (as shown in Fig. 4d).

Depending on annealing temperature (400-700°C), random pores obtained after electrochemical etching become spherical voids. The nearest small voids are then coalesced to form larger voids, as shown in Fig 3. Keeping activation energy, surface atoms of spherical voids jump to form faced voids that are bounded by low-energy facets. The measured angle between the two prominent orientations is in perfect agreement with the angle value of ~54° between the lowest surface energy (100) and (111) oriented facets in crystalline Ge [10],[11].

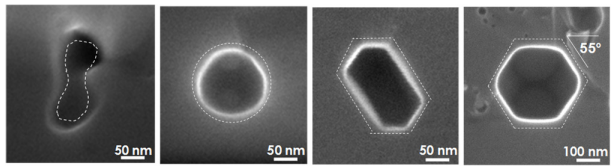

Fig. 3. Different types of the void shapes observed during the annealing of Ge/Si layer (400-700°C) in hydrogen ambient.

*Dislocation-void interactions*

Fig. 4 shows STEM images obtained at the different steps of our process. A dislocation-free region is observed at the upper part of the voided region. Fig. 5 revealed that threading dislocations were terminated at the nanovoid area. Similar effect was observed by Myers et al. he suggests that this overlap is due to an attractive force created by a population of nanovoids with a binding energy that depends upon the surface tension of the void and the relative positions of the interacting dislocation [12]. Also, during thermal annealing, TDs were stopped moving at the void probably because their gliding distance to the voids side-walls were very short [13],[14]. Therefore, the formation of surface-free regions located between the top surface of Ge and Ge/Si interface could be an effective tool to reduce significantly the TDD in Ge/Si layer. The TD free top layers offer a unique potential and a technologically attractive path for overgrowth of high quality Ge based optical devices.

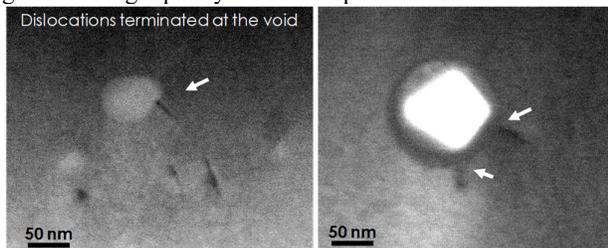

Fig. 5 STEM images shows the termination of a dislocation segment in the nanovoid area.

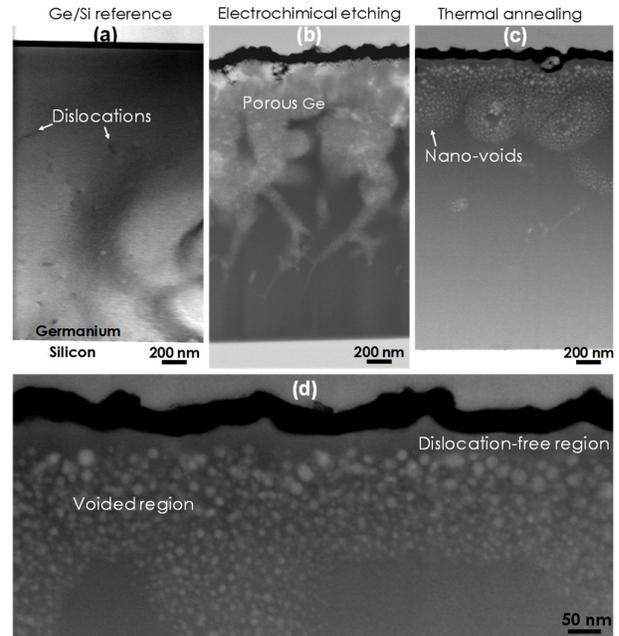

Fig. 4 STEM mixed Dark Field and Bright Field images obtained from irradiated 80 nm Ge/Si sample, (a) reference sample, (b) porous Ge/Si sample, (c) voided Ge/Si sample and (d) higher magnification of (c) indicates the formation of a dislocation-free region at the upper part of the voided region.

### 3. Conclusions

In this work, we proposed a new architecture of low cost virtual substrate engineering. This concept is promising for the fabrication of dislocation-free Ge on Si substrate. The resulting epi-ready surface will be used as the starting point for the deposition of the III–V junctions in the 3J PV cell. Using a simple, inexpensive process through electrochemical etching and thermal annealing will be crucial in cutting manufacturing costs and enabling market penetration of Ge based devices.

**References**
[1] S. Kurtz, *Technical Report*, (2012).
[2] R. R. King, D. C. Law, K. M. Edmondson, C. M. Fetzer, G. S. Kinsey, H. Yoon, R. a. Sherif, and N. H. Karam, *Appl. Phys. Lett.*, 90, 98–100 (2007).
[3] R. Beeler, J. Mathews, C. Weng, J. Tolle, R. Roucka, A. V. G. Chizmeshya, R. Juday, S. Bagchi, J. Menndez, and J. Kouvetakis, *Sol. Energy Mater. Sol. Cells*, 94, 2362–2370, (2010).
[4] H. Ye and J. Yu, *Sci. Technol. Adv. Mater.*, 15, 024601, (2014).
[5] Y. A. Bioud, A. Boucherif, S. Fafard, V. Aimez, R.Ares, *Inter. Conf. Si Epi. heterostructures. procceding, ICSI9*, 130-132, (2015).
[6] Y. A. Bioud, A. Boucherif, A. Belarouci, E. Paradis, D. Drouin, and R. Arès, *Nanoscale Res. Lett.*, 11, 446, (2016).
[7] Y. A. Bioud, A. Boucherif, A. Belarouci, E. Paradis, S. Fafard, V. Aimez, D. Drouin, R. Arès., *Electrochim. Acta*, 232, 422–430 (2017).
[8] M. N. Beattie, Y. A. Bioud, D. G. Hobson, A. Boucherif, C. E. Valdivia, D. Drouin, R. Arès, K. Hinzer, *Nanotechnology*, 29, 215701, (2018).
[9] M. Y. Ghannam, A. S. Alomar, J. Poortmans, and R. P. Mertens, *J. Appl. Phys.* 108, (2010).
[10] S. Tutashkonko, T. Nychyporuk, V. Lysenko, and M. Lemiti, *J. Appl. Phys.*, 113, (2013).
[11] A. Boucherif, G. Beaudin, V. Aimez, and R. Ares, *Appl. Phys. Lett.*, 102, (2013).
[12] S. M Myers and D. M. Follstaedt, *J. Appl. Phys*, 86, 3048–3063, (1999).
[13] M. Raïssi, G. Regula, and J.-L. Lazzari, *Sol. energy Mater. Sol. Cells*, 144, 775, (2016).
[14] M. Yamaguchi, A. Yamamoto, M. Tachikawa, Y. Itoh, and M. Sugo, *Appl. Phys. Lett.*, 53, 2293–2295, (1988).